# IMPROVING BITTORRENT'S PEER SELECTION FOR MULTIMEDIA CONTENT ON-DEMAND DELIVERY

Ananda Görck Streit and Carlo Kleber da Silva Rodrigues

FATECS, University Center of Brasília – UniCEUB, Brasília, DF, Brazil

*ABSTRACT*

*The great efficiency achieved by the BitTorrent protocol for the distribution of large amounts of data inspired its adoption to provide multimedia content on-demand delivery over the Internet. As it is not designed for this purpose, some adjustments have been proposed in order to meet the related QoS requirements like low startup delay and smooth playback continuity. Accordingly, this paper introduces a BitTorrent-like proposal named as Quota-Based Peer Selection (QBPS). This proposal is mainly based on the adaptation of the original peer-selection policy of the BitTorrent protocol. Its validation is achieved by means of simulations and competitive analysis. The final results show that QBPS outperforms other recent proposals of the literature. For instance, it achieves a throughput optimization of up to 48.0% in low-provision capacity scenarios where users are very interactive.*

## 1.INTRODUCTION

The fast growth of companies dedicated to provide multimedia content on-demand motivated studies focusing on maximizing the end-user experience. One potential solution identified by researchers (e.g., [1-5]) is the application of a Peer-to-Peer (P2P) network architecture to supply the demand for high-quality content distribution.

The leading purpose of the above solution is to take advantage of the resources available at the edge of the network, addressing significant or even total autonomy from central servers. Plenty of practical P2P approaches, like the BitTorrent protocol [6], have been successful in the past years. In fact, the great efficiency achieved by BitTorrent for the distribution of large amounts of data inspired its adoption for on-demand video streaming [7-9].

Peer-Selection Policy is the term assigned to one of the core concepts of BitTorrent. By stimulating direct reciprocity and cooperation, its design specifies who makes use of the uplink capacity from all the peers inside the *swarm* (i.e., a group of peers downloading a same content). One observed outcome of this policy is that in a heterogeneous system, i.e., where users have different bandwidth capacities, peers with higher upload capacities typically have higher download speeds than slower peers [7].

However, one essential understanding is that peers receiving a multimedia stream do not need a download rate higher than the playback rate of the media file. Rigorously, all they need is a policy that makes the whole system efficiently allocate its total upload/download capacity so that all of its participants may be served satisfactorily.

Streaming using BitTorrent thus deserves particular concerns focused on the Peer-Selection Policy's concept in order to fulfill QoS requirements like low start up delay and smooth playback continuity. Despite the relevance of the matter, an efficient algorithm capable of equally sharing peers' upload bandwidth among fast and slow users is still open for deeper investigation.

Within this context, this is exactly what this paper is concerned about: to identify a rule to establish the most suitable peers for a user to make content requests so that the whole system dynamics is favored. A novel BitTorrent-like peer-selection strategy named as *Quota-Based Peer Selection* (QBPS) is introduced herein with the goal of bringing new insights into this promising field of study.

The novel proposal is based on a quota-assignment policy, where peers experiencing lower download rates get more opportunities to access the content on time to be streamed. The idea is to promote high QoS for on-demand video streaming by efficiently balancing peers' real bandwidth requirements as the proposal pursues to provide equal bartering opportunities to all of them.

The validation of this novel peer-selection strategy is achieved by means of simulations and competitive analysis. Moreover, to give a feedback closer to a real VoD system, these simulations are carried out under interactive scenarios, where users are allowed to make interactive requests such as *Play*, *Pause*, *Resume*, *Jump Forwards* and *Jump Backwards*. The final results indicate significant optimizations. For example, the QBPS proposal has shown a throughput optimization of up to 48.0%, compared to other proposals of the literature, in scenarios with low-provision capacity where users are very interactive.

The rest of this work is organized as follows. Section 2 describes the main rules applied by the original BitTorrent system. Next, in Section 3, related work regarding the adaptation of BitTorrent to VoD systems is discussed and analyzed. It is in Section 4 that the new proposal presented by this paper is introduced. The results and corresponding analysis lie in Section 5. Finally, conclusions and future work guidance are found in Section 6.

## 2. BASIS OF A BITTORRENT SYSTEM

The purpose of the BitTorrent protocol's design is to provide a pretty simple and very efficient content distribution mechanism over the Internet even when the number of participants increases at an unbounded rate. Two main policies are the basis of the resulting efficiency achieved by BitTorrent.

The Piece-Selection Policy determines which are the portions of the content that are selected to be exchanged among the peers. These portions are pieces with a size of typically 256.0 kB. Subsequently, each piece is split into blocks of 16.0 kB in size. These blocks represent the transmission unit used on the piece exchange among the peers.

The selection of the pieces mainly depends on the number of replicated units of each piece the system already has available. More precisely, the probability a piece is selected is higher when it is rare and, consequently, is one of the least replicated in the swarm in comparison to the other pieces of the file. The observed number of replications though considers the local view of an arbitrary peer and is subjected to the pieces already obtained by the remote peers connected to the local peer.

The other policy, named as Peer Selection, determines with which of the connected remote peers a local peer distributes his content. BitTorrent promotes reciprocation by determining that at every 10 seconds peers regularly *unchoke* (i.e., enable a peer to request data) other peers who have given them needed content with the highest speed rates. A local peer can only choose for *unchoke* remote peers that are *interested* in him, i.e., remote peers that do not have a piece the local peer has.

By default, the protocol equally divides the upload bandwidth into four upload slots, three of them (called regular slots) directed to perform reciprocation. The fourth one (called optimistic slot) is used to promote altruism by randomly selecting, at every 30 seconds, a peer from all the interested remote peers the local peer is connected to. This upload slot is used (1) to find other peers with a higher transmission rate and also (2) to bootstrap new peers by helping them to get their first pieces.

BitTorrent also defines a centralized entity named tracker. It is responsible to send a list of random peers inside the swarm to new users. It is with this list that a user can make connections to other peers interested in downloading the same content as him. By default, the protocol specifies that a peer may have a minimum of 20 and a maximum of 80 connected remote peers. Only 40 connection invitations may be though sent by the user, opening space for him to accept 40 more invitations sent by other peers [8].

When a peer finishes his download (i.e., he has all the pieces of the content being distributed) he changes from *leecher* state to *seeder* state. The leecher's role is to receive content while, at the same time, he contributes with his own resources to the swarm. The reciprocation scheme introduced by the Peer-Selection policy is what motivates leechers to share their own resources. By its turn, seeders work exclusively as servers and consequently BitTorrent's incentive mechanisms cannot reward their upload bandwidth contributions.

## 3. RELATED WORK ON ADAPTING BITTORRENT

Prior works mostly concentrated their efforts on characterizing and analyzing the adjustment of the Piece-Selection Policy for media streaming [7, 9-16]. Only more recent works [3, 17-19] target the adaptation of the Peer-Selection Policy for streaming. They basically alter the following major characteristics of the algorithm: (1) quantity of upload slots a peer opens; (2) leecher's and seeder's rule to decide the preferred peers for upload; and (3) time-cycle for re-evaluation of the slots.

To define these characteristics, the subsequent points are important to consider: how the content's provider can rapidly spread parts of the file; how altruistic a peer should be without compromising his performance; how to promote cooperation and still fully utilize resources and balance them in the swarm; and by what means newcomers receive their first pieces.

The Give-to-Get (G2G) protocol [20] is one example regarding modifications on the characteristic (1) of the original algorithm. It designates one optimistic upload slot and a minimum of three regular upload slots as it occurs in BitTorrent. There may be more than three slots though (a maximum of two more), if there is still spare upload capacity left due to current downloading bottlenecks on the already unchoked peers. The attempt is to maximize the bandwidth utilization by increasing the means of cooperation among peers.

The characteristic (1) is also focused in the work of D'Acunto et al. [7]. Their proposal is mainly devoted to heterogeneous environments and, accordingly, the number of upload slots is adjustable depending on the peer's total upload bandwidth. They design three different schemes under this scenario. The idea of the first scheme is to share the bandwidth of high-capacity users with more peers, limiting the rate provided by each slot to the playback rate of the media file and thus augmenting cooperation among peers. The second scheme defines that peers should be more altruistic by dynamically adjusting the number of their optimistic unchoke slots to their current QoS. Finally, the third scheme is related to the characteristic (2) as it prioritizes newcomers when performing optimistic unchokes, reducing the delay new peers may experience before starting the playback of the content.

Changes on the characteristic (2) also appear in the work of Rocha & Rodrigues [21]. The rule for selecting peers prioritizes those presenting the lowest possible dispersion. To do so, the following three values ought to be collected: (a) how much the piece requests diverge from each other considering the arrival times; (b) how much of the retrieved file segments may be effectively shared; and (3) how often each file position is requested. They also analyze different interactive users' profiles, demonstrating that highly interactive users tend to present greater dispersion. By minimizing the dispersion (temporal and local), peers can better exploit their uplink capacities by exchanging pieces that interest them.

Wen et al. [22] proposes the selection of peers with the closest playback point and consequently with a lower local dispersion. This rule encourages the mutual cooperation among peers with common interests. The work of Carlsson et al. [18] proposes another algorithm that changes specifically the seeder's rule of characteristic (2). They define that seeders should preferentially allocate server bandwidth to send rare pieces to newly arrived peers and to peers at imminent risk of receiving data too late for playback.

Finally, three proposals developed by Rodrigues [8] are explained next. One of them intends to equally share the total upload bandwidth among regular and optimistic slots, respectively, clearly changing the characteristic (2) of the original algorithm. The other proposal aims at providing more opportunities for newcomers by modifying the characteristics (2) and (3), respectively. It defines that peers should have three optimistic upload slots and only one regular slot. The time interval for re-evaluation of the optimistic slots decreases 1/3 in comparison to the original setup. The third proposal modifies only the characteristic (2) by prioritizing for unchoke the fastest peers that are waiting longer to receive a piece. All proposals determine that seeders should select remote peers that have been unchoked more recently.

It is worth mentioning that interactivity brings different features to weight when shaping BitTorrent for working in VoD systems. As mentioned by Rocha & Rodrigues [21], high interactive users tend to stay less time inside the swarm and consequently are likely to requests less data than peers streaming in a sequential fashion. Also, the segments of data requested are mostly dispersed over the media file, provoking a higher number of interruptions and therefore spending more time on slots for data download.

Moreover, if the reader is interested, the studies of D'Acunto et al. [9] and Ma et al. [23] debate and examine differences achieved in the overall system performance when distinct piece-selection and peer-selection strategies are composed and tested in conjunction [15]. It is important to have this in mind in Section 5, where the new peer selection proposed herein (described in Section 4) is simulated together with a different piece-selection algorithm instead of the original one from BitTorrent.

## 4. QUOTA-BASED PEER SELECTION

This section presents the novel BitTorrent-like protocol denoted as *Quota-Based Peer Selection* (QBPS). For ease of presentation and objectivity, only the parts modified from the original protocol are described herein. As already mentioned, the proposal is based on a quota-assignment policy, where peers experiencing lower download rates get more opportunities to access the content on time to be streamed.

To do so, leechers divide their upload bandwidth into four different data upload slots. A variable named $MAX_{QUOTA}$ specifies the maximum number of upload slots destined to improve the altruism of peers. This variable depends on the rates of the local peer and of his interested peers, respectively.

Leechers only allocate quota slots to remote peers that have a download rate slower than their own rates. If the local peer has a high download rate, then more remote peers are candidates to occupy a quota slot. The whole system therefore tends to balance its uplink capacity, endorsing peers to collaborate with others holding scarcer bandwidth.

Peers with common interests are stimulated to mutually upload closely located pieces (i.e., with lower dispersion). The buffer tends to cover an area that will be soon played back by the users, close to the reproduction (i.e., current playback) point. Thus, quota-slot candidates are classified according to their playback points; those watching closer parts of the media file are prioritized for selection.

The idea is to promote an altruistic posture for leechers that have a high sharing capability and still ease the discovery of peers intending to download similar parts of the file. Note that remote peers that have not exchanged pieces with the local peer yet are considered to have rates equal to zero and therefore are candidates to occupy quota slots. As a matter of fact, the sending peer can only determine the rates of the receiving peers when pieces are exchanged. If the rate of a new peer is high, he can easily reciprocate. On the other hand, if the rate is low, he is very likely to soon receive pieces from others holding better download performances.

The rest of the upload slots not destined to apply quotas work identically as the regular slots from the original BitTorrent protocol. By doing so, QBPS likewise incites reciprocation and cooperation among peers. Moreover, a seeder operates the same way as in the original BitTorrent: three slots are offered to the fastest downloading peers connected to him and one is designated for the discovery of new fastest leechers. In the end, data flows from the provider to the fastest leechers and later to newcomers or leechers with lower data throughput.

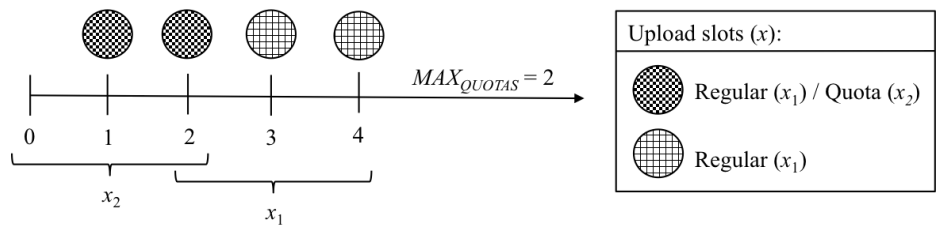

Figure 1. Upload slots of leechers under the QBPS proposal

Figure 1 exemplifies how the division of upload slots works for leechers. They have a total of $x = x_1 + x_2$ upload slots, where $x_1$ represents regular slots and $x_2$ represents quota slots. In the example, $MAX_{QUOTA} = 2$, resulting in an interval where there are $x_1 = 2, \ldots, 4$ regular slots and an interval where there are $x_2 = 0, \ldots, 2$ quota slots.

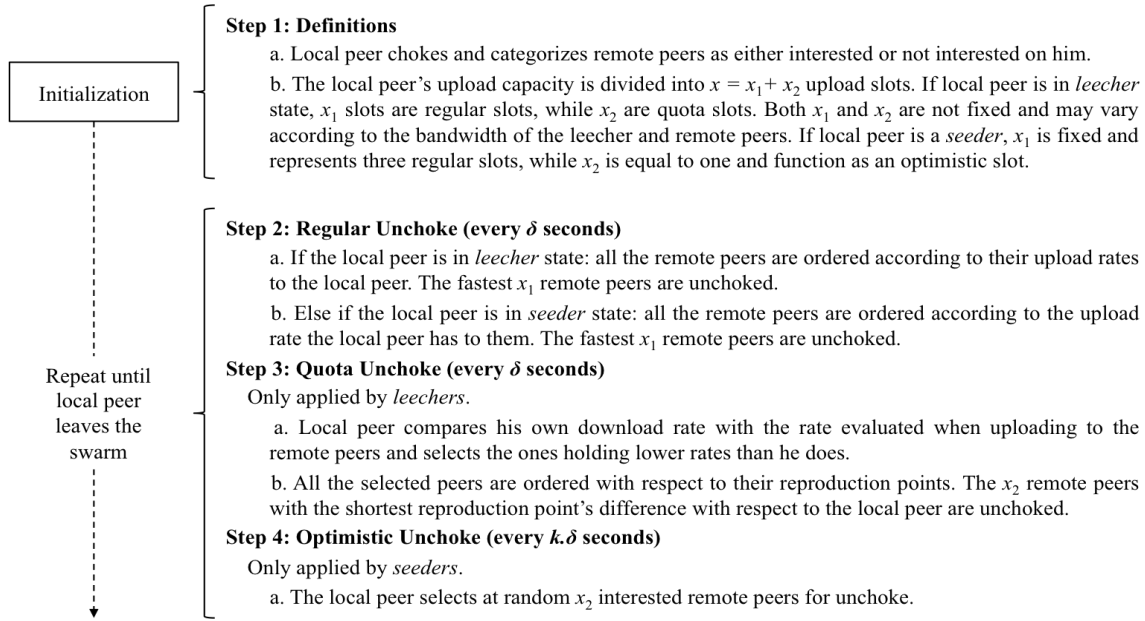

Figure 2. Overall operation of QBPS.

Figure 2 is a general guideline for understanding the operation of the QBPS protocol. The amount of slots a peer opens for data upload is represented by *x*. Its value is fixed, but the variables $x_1$ and $x_2$ may change depending on the value assigned to $MAX_{QUOTA}$ and in which state the local peer is.

## 5. PERFORMANCE EVALUATION

This section is divided into three subsections. Subsection 5.1 explains the scenario characterization used in the simulations, especially outlining the past works that serve as a guideline for this purpose. Subsection 5.2 focuses on defining the simulation setup and the performance metrics used to evaluate the proposals examined herein. Lastly, Subsection 5.3 presents the results and the corresponding analysis. Note that the simulations shown by Streit & Rodrigues [15] employed a quite similar scenario configuration.

### 5.1. SCENARIO CHARACTERIZATION AND WORKLOADS

Several works in the literature (e.g., [24-28]) analyze real streaming multimedia workloads to establish realistic simulation scenarios. These scenarios are organized with different parameters and functionalities as shown in Table 1. The choice of each parameter value is conditioned to a general media-streaming scenario and is based on several works of the literature such as [7, 27, 29-32].

Table 1. Scenario characterization parameters for a general media-streaming scenario.

| Aspect | Symbol | Definition | Value | | |
|---|---|---|---|---|---|
| Content | $O_s$ | Content size, measured in bytes | 20.0 MB | | |
| | $p_s$ | Piece size, measured in bytes | 256.0 kB | | |
| | $b_s$ | Block size, measured in bytes | 16.0 kB | | |
| | $R$ | Reproduction rate, measured in bits per second | 240.0 kbps | | |
| Interactivity | $I_p$ | Interactive profile | HI | MI | LI |
| | $d_0$ | Mean exponential time in state Play, in seconds | 1.20 | 1.70 | 2.20 |
| | $d_1$ | Mean exponential time in state Stop, in seconds | 0.00 | 0.00 | 0.00 |
| | $d_2$ | Mean exponential time in state Pause, in seconds | 1.00 | 1.00 | 1.00 |
| | $d_3$ | Mean exponential time in state JB, in seconds | 0.75 | 0.75 | 0.75 |
| | $d_4$ | Mean exponential time in state JF, in seconds | 0.75 | 0.75 | 0.75 |
| | $p_0$ | Transition probability from Play to Play | 0.35 | 0.60 | 0.85 |
| | $p_1$ | Transition probability from Play to Stop | 0.05 | 0.04 | 0.02 |
| | $p_2$ | Transition probability from Play to Pause | 0.20 | 0.12 | 0.04 |
| | $p_3$ | Transition probability from Play to JB | 0.20 | 0.12 | 0.04 |
| | $p_4$ | Transition probability from Play to JF | 0.20 | 0.12 | 0.04 |
| BitTorrent swarm | $n$ | Number of seeders in the swarm | 1 | | |
| | $m$ | Number of leechers in the swarm | 20 | | |
| | $P_c$ | Average scenario provision capacity | OP | LP | BP |
| | | | 1.25$R$ | 0.8$R$ | 1.0$R$ |
| | $S_{down/up}$ | Download and upload rate of seeders, measured in bits per second | 240.0 kbps | | |
| | $L_{down/up}$ | Download and upload rate of leechers, measured in bits per second | High | Low | Regular |
| | | | 480.0 kbps | 120.0 kbps | 240.0 kbps |

Regarding the scenario provision capacity, there are three categories [7], which are defined as it follows. The first category is denoted as *Over-Provision* (OP). This refers to a heterogeneous scenario where 50% of the peers possess high capacity (2$R$), and the other half has slow capacity (0.5$R$), resulting in a system with average capacity equal to 1.25$R$. The second category is denoted as *Low-Provision* (LP). This is a heterogeneous scenario with 20% of the peers having high capacity (2$R$), and the left 80% having low capacity (0.5$R$), resulting in a system with an average capacity equal to 0.8$R$. At last, the third category is denoted as *Balanced-Provision* (BP) and refers to a homogeneous scenario where 100% of the peers have a download/upload capacity equal to $R$.

With respect to the user's interactivity, the media reproduction always begins with a *Play* interaction and continues up to the end of the media or up to a *Stop* interaction. While streaming, users can perform intermediate actions like *Play*, *Stop*, *Pause*, *Jump Forwards* (JF) and *Jump Backwards* (JB). To emulate the user's interactive actions, synthetic workloads are generated by a user interactive model [33] and three different interactive profiles are then devised: *High Interactivity* (HI), *Medium Interactivity* (MI) and *Low Interactivity* (LI). At last, the interactive model has five different states, each of them corresponding to one interactive action. The duration of each state is defined by an exponential distribution of mean $d_i$, $i = 0, \ldots, 4$, while the transitions between the states occur with probabilities $p_i$, $i = 0, \ldots, 4$.

### 5.2. SIMULATION SETUP AND PERFORMANCE METRICS

All proposals examined herein are modeled on top of the Tangram-II modeling environment [34], which is an event-driven object-oriented simulation tool. It is worth noting that in order to evaluate a BitTorrent-protocol we need to consider the modeling of both a piece-selection policy and a peer-selection policy. With this in mind, we explain the simulation setup.

For the peer-selection policy, we consider: the QBPS policy, which is the novel proposal introduced in this work; the *Select Balanced Neighbour Policy* (SBNP) [8], which is a recent and efficient proposal of the literature and so it properly serves for the competitive analysis we carry out in what follows; and the original BitTorrent's peer-selection policy, which sets a threshold to evaluate the gained optimization provided by the novel proposal.

For the piece-selection policy, the *Adaptive-Definite Window Interactive Streaming* (ADWIS) [15] is chosen. This choice mainly lies on the fact that it has already proven to be one of the most efficient proposal for piece-selection in the literature and, after all, it does not really influence on the competitive analysis to be carried out later in this text since it is implemented the same way in all the three proposals for peer-selection previously mentioned.

Table 2. Performance metrics.

| Metric | Notation | Definition |
|---|---|---|
| Efficiency Retrieving Coefficient | ERC | Evaluates the download efficiency achieved by the peers in comparison to a situation where the file retrieving process occurs under an exclusive data-delivery channel, when the peers do not suffer from playback interruptions. |
| Number of Peers Served | PS | Mainly used to see whether an expressive number of peers have been served during the whole simulation. |
| Mean Empty Slot Time | EST | Average time peers spend without distributing resources because their upload slots are empty, considering all the peers that participated in the swarm. |
| Mean Startup Delay | SD | Average time spent before peers start playing back, considering all the peers that participated in the swarm. |
| Mean Number of Interruptions | NI | Average number of missing data pieces occurred in a simulation run, considering all the peers that participated in the swarm. |
| Mean Time to Return | TR | Average delay time suffered by all the peers in a simulation run after an interruption. |

It is worth saying that the ADWIS proposal introduces a playback window set ($w_{adwis}$) which covers pieces with higher download priority and adapts its size accordingly to the peer's download performance. Besides, it also defines a threshold $\theta$ that places a lower bound on the number of contiguous pieces required before the window can grow. At last, although this proposal has another window defined for the piece-selection mechanism, for the sake of simplicity, only the reproduction window is implemented in this work [15].

The whole system considered in the experiments is always in steady state. This means that, even when peers join and leave the system, the number of active peers, represented by the sum of *m* and *n*, respectively, stays constant. Hence, the total number of peers served in a simulation run does not correspond to the total number of concurrent active peers. The total number of peers served is actually a performance metric and it is described together with the other metrics given in Table 2. Furthermore, the simulation results have confidence intervals of 95.0%, which are within 5.0% around the average of the metric values presented herein.

Table 3. Parameter values of each policy modeled.

| Peer Selection | | | | | | | | Piece Selection | |
|---|---|---|---|---|---|---|---|---|---|
| QBPS | | | | SBNP [8] | | | | ADWIS [15] | |
| $MAX_{QUOTA}$ | $x$ | $k$ | $\delta$ | $x_1$ | $x_2$ | $k$ | $\delta$ | $w_{adwis}$ | $\theta$ |
| 2 | 4 | 3 | 10 | 2 | 2 | 3 | 10 | 7 | 3 |

Lastly, Table 3 describes numerical values used by each of the proposals. Most of them are extracted from past literature works [8, 15] as well as from numerous experiments carried out while elaborating this work. These experiments are not illustrated herein due to the sake of objectivity.

### 5.3. RESULTS AND COMPETITIVE ANALYSIS

#### 5.3.1. SYSTEM DYNAMICS ANALYSIS

This subsection has the prior goal to competitively understand the system performance achieved when applying the novel proposal QBPS, the SBNP proposal and the original peer-selection policy of BitTorrent. For this analysis, the following performance metrics are evaluated: *Efficiency Retrieving Coefficient* (ERC), *Number of Peers Served* (PS) and *Mean Empty Slot Time* (EST).

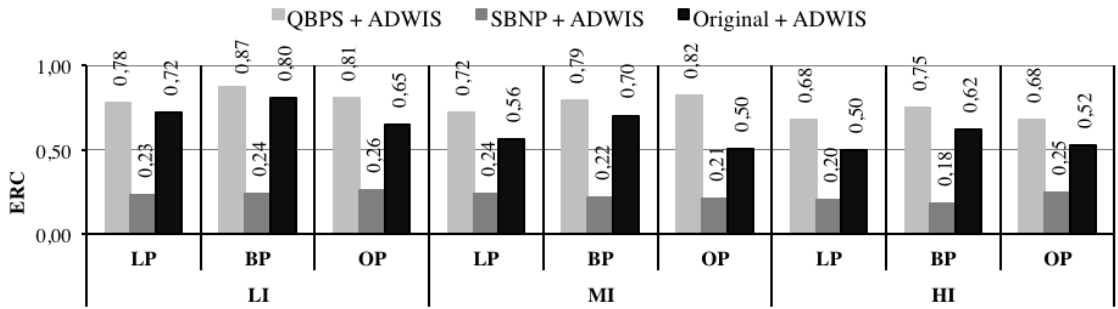

Figure 3: ERC for distinct policies

The values obtained for the metric ERC are depicted in Figure 3. Higher values indicate a better efficiency for the file retrieving process. As it can be observed, QBPS outperforms the other policies, no matter the scenario and the interactive level considered. For instance, the results have shown an improvement of about 18.0% compared to the original peer-selection policy of BitTorrent, and of about 48.0% compared to the SBNP proposal in scenarios with low-provision capacity where users are very interactive. This definitely indicates that under this policy more peers efficiently download pieces of the media file. Even if SBNP presents a more consistent ERC under distinct conditions, showing a greater balance than the other policies, it does not enhances the system throughput. Its performance is even lower than the one achieved by the original peer-selection policy of BitTorrent.

The total number of peers served in each swarm session is presented in Figure 4. When the peer-selection strategy promotes a good balance between altruism and reciprocation, it provides to all the active peers a higher throughput, a higher ERC and consequently a more efficient piece exchange in the system. This results in more users being able to participate in the swarm during the same amount of time. As shown in the same figure, the novel proposal QBPS has higher values for the metric PS compared to the other policies, even in low-provision scenarios. Thus, this exhibits that it can best maintain the needed efficiency for all the peers joining streaming BitTorrent-like systems.

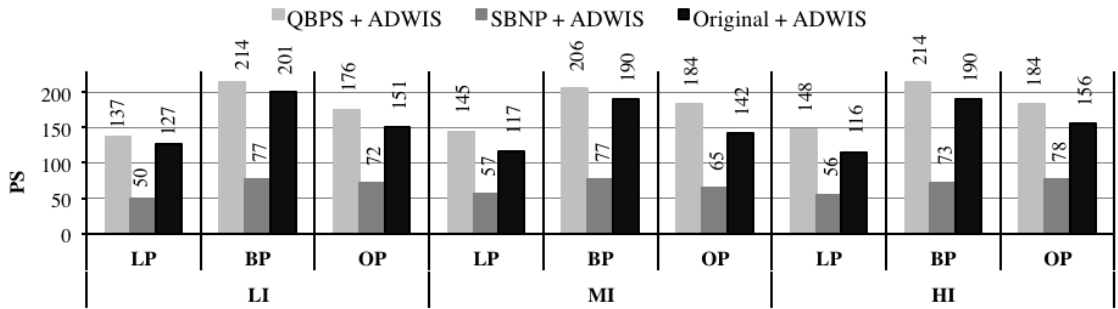

Figure 4: PS for distinct policies

Finally, the results observed for the metric EST are in Figure 5. Small values for this metric are positive and occur when peers do not stop sharing their pieces to the system. The SBNP proposal has the worst results. This can be explained by the reduction of cooperation incentives to peers in order to increase their altruistic nature. Less cooperation among peers stands for QoS degradation, especially when the bandwidth is scarce, as peers are not stimulated to mutually share and enhance their piece retrieving progress. QBPS also diminishes reciprocation for the same reason, but this happens only when users are in a better download situation than their remote peers.

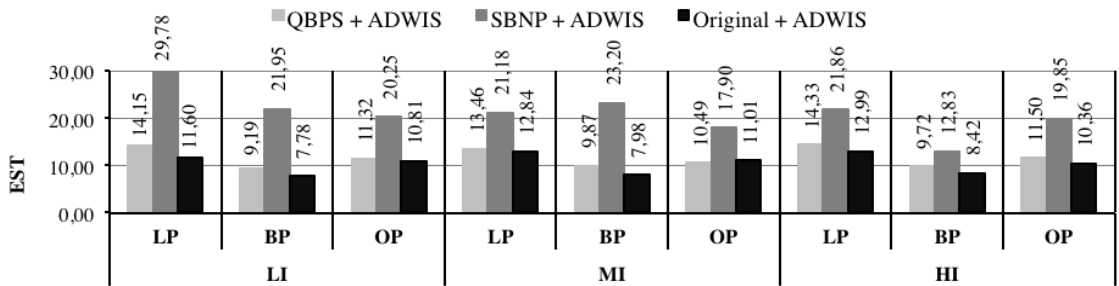

Figure 5: EST for distinct policies

Moreover, compared to the original peer-selection policy, the QBPS proposal maintains upload slots unutilized for more time. This specifically shows that this proposal has some space for improvements, as the peers under its strategy could contribute with more of their uplink capacity to share resources. On the other hand, this proposal has showed better results for all the other metrics then observed above. Accordingly, even if QBPS could promote a higher utilization of the peer's sharing capability, the slots are clearly better utilized and so its selection policy can be seen as the most efficient one.

### 5.3.2. USER EXPERIENCE ANALYSIS

This subsection reports the results obtained from the user experience point of view. This investigation lies specifically on the following performance metrics: *Mean Startup Delay* (SD), *Mean Number of Interruptions* (NI) and *Mean Time to Return* (TR). Better results are achieved when all these metrics are reduced. Also, note that the results demonstrated in the last subsection highly influence the results presented herein.

Figure 6 plots the outcomes for the metric SD. Small differences are achieved among the interactive profiles. For distinct system's average capacity, the homogeneous scenario (BP) presents the best results for this metric in most of the proposals. In heterogeneous scenarios (LP and OP), the metric is clearly reduced the higher the system's average capacity is.

Observe that better results are obtained when adopting the original and the QBPS peer-selection policies. Even though SBNP equally divides the peer's upload slots between optimistic and regular slots, increasing the probability of leechers to select new peers, seeders lack of optimistic unchokes. Newcomers therefore uniquely depend on leechers for collecting the first pieces of the media file, augmenting the startup delay especially in heterogeneous scenarios, where the rates provided for data download might be the half of the reproduction rate.

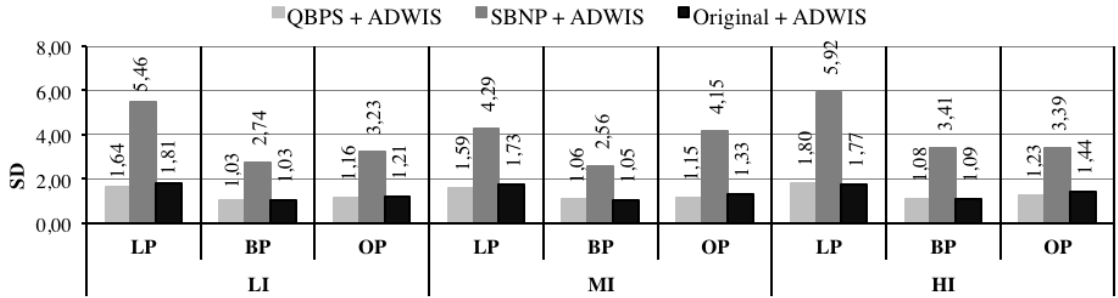

Figure 6: SD for distinct policies

Comparing the values obtained for NI in Figure 7, QBPS presents better results in most of the scenarios. For instance, this proposal exceeds its performance in over-provision scenarios, demonstrating that the use of quota slots (mostly by leechers with greater rates) is a good strategy for balancing the system and, consequently, for providing to all peers a higher number of pieces before their deadlines are reached. Although the optimistic slots implemented by the original and the SBNP policies are also a tool utilized for balancing the system, its rule determines the selection of random interested peers and does not specifically focus on granting pieces for peers with impaired download rates.

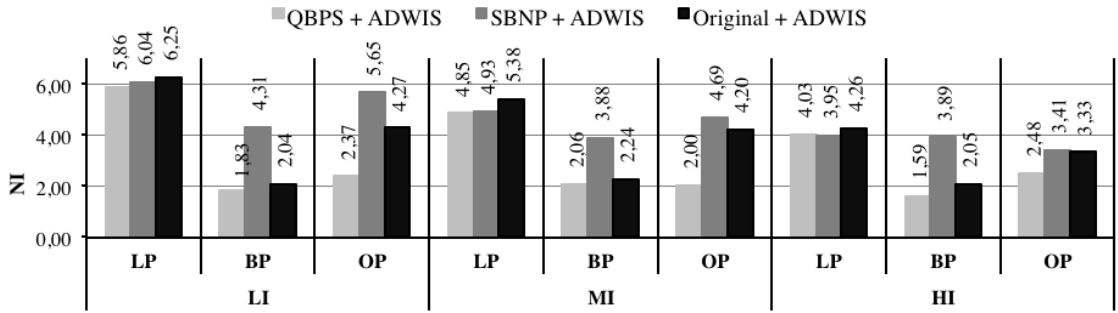

Figure 7: NI for distinct policies

In low-provision scenarios, the performances are similar for all the proposals. Even when the strategies of them intend to balance the upload among peers, it is still a challenge to reduce the number of interruptions when the total upload capacity of the system is smaller than the playback rate. Again, the results from the SBNP have a bigger disparity comparing to the other proposals. In fact, the high altruism proposed by this strategy, even for peers with small downloading rates, might be the reason for its poor achievement, specially intensified in low-provision scenarios.

Furthermore, note that high-interactive users tend to suffer fewer interruptions independently of the system average capacity. This is basically due to the frequency in which some pieces are watched more than once. For example, when jumping, peers may go to parts of the file that have already been reproduced (and downloaded), increasing their time to recover a more diverse set of pieces without compromising the user's experience.

As for the metric TR, indicated in Figure 8, QBPS attains the best results. This indicates that, even when a higher number of interruptions is suffered by the users, specially in LP scenarios as shown in Figure 4, the quota slots are the most successful mechanism to reduce the amount of time a peer holding a lower download rate has to wait before the playback returns. The strategies from the other policies used to promote altruism evidently harm the user's waiting time to continue the playback as they do not focus on granting uplink capacity to interrupted users (normally the ones possessing lower download rates).

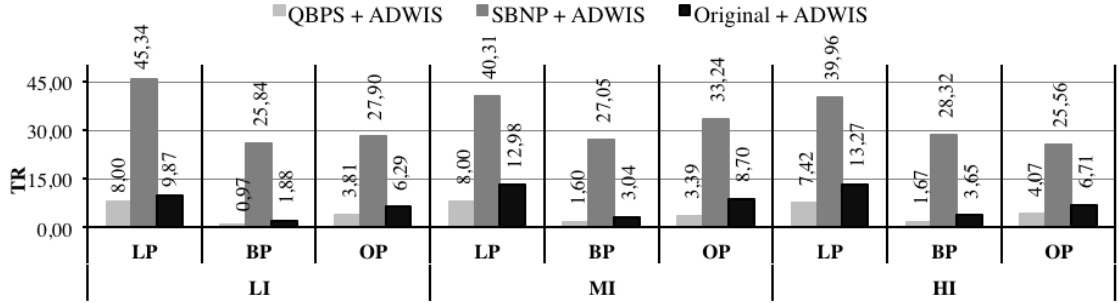

Figure 8: TR for distinct policies

Considering the above, the employment of the QBPS policy is the most appropriate one to be considered for providing a better user's experience to all the peers participating in a swarm session. Altruism is noted to be a really important tool to be applied in BitTorrent-like streaming services. However, it must be carefully employed to really benefit the system. In fact, the quota slots from QBPS are seen to be a strategy that positively promotes a balanced altruism among peers, focusing on diminishing bandwidth inequalities as they target at the selection of peers suffering from low download rates and therefore more susceptible to have their playback interrupted.

## 6. CONCLUSIONS AND FUTURE WORK

This work presented a novel BitTorrent-like peer-selection strategy for interactive multimedia streaming: the *Quota-Based Peer Selection* (QBPS). The proposal is based on a quota-assignment policy, where peers experiencing lower download rates get more opportunities to access the content on time to be streamed. The idea is to promote high QoS for on-demand video streaming by efficiently balancing peers' real bandwidth requirements as the proposal pursues to provide equal bartering opportunities to all of them.

Two important conclusions may be highlighted among those obtained in this work. First, the QBPS proposal showed to be very competitive compared to the original peer-selection policy of BitTorrent and to another state-of-art peer-selection policy from the literature. It outperforms both in most of the simulations carried out in this work, showing that, not only the system dynamics is enhanced and better utilized when employing it, but also that the user achieves a smoother playback experience. For example, the results have shown a throughput optimization of up to 48.0% in scenarios with low-provision capacity where users are very interactive. This indicates that employing quota slots is a really interesting strategy, as it promotes a balanced altruism among peers when focusing on diminishing bandwidth inequalities.

Second, altruism is an important design characteristic to be applied under interactive streaming scenarios, especially in heterogeneous systems. However, the optimistic slots strategy, originally proposed by the BitTorrent's peer-selection policy, is not an optimal solution and should be replaced by another that only stimulates an altruistic posture to peers possessing higher download capabilities, as it is done by the quota slots of the QBPS policy, proposed in this work.

Finally, future work may include further analysis in the application of quota slots to better understand its properties and operation, including the deployment of alternative rules to determine which peers holding lower download rates should be elected for data upload using these slots.

**REFERENCES**

[1] Huang, Y., Fu, T. Z., Chiu, D. M., Lui, J., & Huang, C. (2008) "Challenges, design and analysis of a large-scale p2p-vod system", In ACM SIGCOMM Computer Communication Review, Vol. 38, No. 4, pp 375 – 388
[2] Hammami, C., Gazdar, A., Jemili, I., & Belghith, A. (2015) "Study of VOD streaming on BitTorrent", In Networks, Computers and Communications (ISNCC), pp 1 – 6.
[3] Yang, Y., Chow, A. L., Golubchik, L., & Bragg, D. (2010), "Improving QoS in bittorrent-like VoD systems", In INFOCOM 2010 Proceedings IEEE, pp 1 – 9.
[4] Liang, C., Fu, Z., Liu, Y., & Wu, C. W. (2010), "Incentivized peer-assisted streaming for on-demand services", Parallel and Distributed Systems, Vol. 21, No. 9, pp 1354 – 1367.
[5] Shen, Z., Luo, J., Zimmermann, R., & Vasilakos, A. V. (2011), "Peer-to-peer media streaming: Insights and new developments", Proceedings of the IEEE, Vol. 99, No. 12, pp 2089 – 2109.
[6] B. Cohen (2003), "Incentives Build Robustness in BitTorrent", In Workshop on Economics of Peer-to- Peer Systems, Vol. 6, pp 68 – 72.
[7] D'Acunto, L., Andrade, N., Pouwelse, J., & Sips, H. (2010), "Peer selection strategies for improved QoS in heterogeneous BitTorrent-like VoD systems", In Multimedia (ISM), pp 89 – 96.
[8] Rodrigues, C. K. D. S. (2014), "Analyzing peer selection policies for BitTorrent multimedia on-demand streaming systems in internet", International Journal of Computer Networks & Communications (IJCNC), Vol. 6, No. 1, pp 203 – 221.
[9] D'Acunto, L., Chiluka, N., Vinkó, T., & Sips, H. (2013), "BitTorrent-like P2P approaches for VoD: A comparative study", Computer Networks, Vol. 57, No. 5, pp 1253 – 1276.
[10] Vlavianos, A., Iliofotou, M. & Faloutsos, M. (2006), "BiToS: Enhancing BitTorrent for Supporting Streaming Applications", In 9th IEEE Global Internet Symposium, pp 1 – 6.
[11] Shah, P. & Pâris, J.-F. (2007), "Peer-to-Peer Multimedia Streaming using BitTorrent", In IEEE International Performance, Computing, and Communications Conference – IPCCC, pp 340 – 347.
[12] Savolainen, P., Raatikainen, N. & Tarkome, S. (2008), "Windowing BitTorrent for Video-on-Demand: Not all is Lost with Tit-for-Tat", In IEEE GLOBECOM, pp 1 – 6.
[13] Borghol, Y., Ardon, S., Carlsson, N. & Mahanti, A. (2010), "Toward Efficient On-Demand Streaming with Bittorrent", In IFIP Networking, pp 53 – 66.
[14] Streit, A. G. & Rodrigues, C.K.S. (2013), "On the Design of Protocols for Efficient Multimedia Streaming over Internet", ESPE - Ciencia y Tecnología, Vol. 4, No. 1, pp 25 – 39.
[15] Streit, A. G., & Rodrigues, C. K. S. (2015), "Proposing a BitTorrent-Like Protocol for Efficient Interactive Multimedia Streaming Applications", The SIJ Transactions on Computer Networks & Communication Engineering (CNCE), Vol. 3, No. 2, pp 28 – 37.
[16] Lin, C. S., & Lin, J. W. (2015), "UR-aware: Streaming videos over BitTorrent with balanced playback urgency and rareness distribution", Peer-to-Peer Networking and Applications, pp 1 – 12.
[17] Rodrigues, C.K.S. (2014), "On the Optimization of BitTorrent-Like Protocols for Interactive On-Demand Streaming Systems", International Journal of Computer Networks & Communications, Vol. 6, No. 5,pp 39 – 58.
[18] Carlsson, N., Eager, D. L., & Mahanti, A. (2009), "Peer-assisted on-demand video streaming with selfish peers", In NETWORKING 2009, pp 586 – 599.
[19] Chow, A. L., Golubchik, L., & Misra, V. (2008), "Improving BitTorrent: a simple approach", In IPTPS, pp 8.
[20] Mol, J., Pouwelse, J., Meulpolder, M., Epema, D. & Sips, H. (2008), "Give-to-Get: Free-riding-resilient video-on-demand in P2P systems", SPIE MMCN, San Jose, California, USA.
[21] Rocha, M. V., & Rodrigues, C. K. S. (2013), "On client interactive behaviour to design peer selection policies for BitTorrent-like protocols", International Journal of Computer Networks & Communications (IJCNC), Vol.5, No.5, pp 141 – 159.
[22] Wen, Z., Liu, N., Yeung, K. L., & Lei, Z. (2011), "Closest playback-point first: A new peer selection algorithm for P2P VoD systems", In Global Telecommunications Conference (GLOBECOM 2011), pp 1 – 5.

[23] Ma, Z., Xu, K. & Zhong, Y. (2012), "Exploring the Policy Selection of P2P VoD System: a Simulation Based Research", In Proceedings of the 2012 IEEE 20th International Workshop on Quality of Service, pp 23.
[24] Costa, C. P., Cunha, I. S., Borges, A., Ramos, C. V., Rocha, M. M., Almeida, J. M. & Ribeiro-Neto, B. (2004), "Analyzing Client Interactivity in Streaming Media", In Proceedings of the 13th International Conference on World Wide Web, pp 534 – 543
[25] Guo, L., Chen, S., Xiao, Z., Tan, E., Ding, X. & Zhang, X. (2005), "Measurements, Analysis, and Modeling of BitTorrent-Like Systems", In Proceedings of the 5th ACM SIGCOMM conference on Internet Measurement, pp 4.
[26] García, R., Pañeda, X. G., García, V., Melendi, D. & Vilas, M. (2007), "Statistical Characterization of a Real Video on Demand Service: User Behaviour and Streaming-Media Workload Analysis", Simulation Modelling Practice and Theory, Vol. 15, No. 6, pp 672 – 689.
[27] Zhang, C., Dhungel, P., Wu, D. & Ross, K. W. (2011), "Unraveling the Bittorrent Ecosystem", Parallel and Distributed Systems, Vol. 22, No. 7, pp 1164 – 1177.
[28] Chen, Y., Zhang, B., Liu, Y. & Zhu, W. (2013), "Measurement and Modeling of Video Watching Time in a Large-Scale Internet Video-on-Demand System", Multimedia, IEEE Transactions, Vol. 15, No. 8, pp 2087 – 2098.
[29] Rocha, M., Maia, M., Cunha, Í., Almeida, J. & Campos, S. (2005), "Scalable Media Streaming to Interactive Users", In Proceedings of the 13th Annual ACM International Conference on Multimedia, pp 966 – 975.
[30] Hoßfeld, T., Lehrieder, F., Hock, D., Oechsner, S., Despotovic, Z., Kellerer, W. & Michel, M. (2011), "Characterization of BitTorrent Swarms and their Distribution in the Internet", Computer Networks, Vol. 55, No. 5, pp 1197 – 1215.
[31] Wang, H., Liu, J. & Xu, K. (2012), "Understand Traffic Locality of Peer-to-Peer Video File Swarming", Computer Communications, Vol. 35, No. 15, pp 1930 – 1937.
[32] Xia, R. L. & Muppala, J. K. (2010), "A Survey of BitTorrent Performance", IEEE Communications Surveys & Tutorials, Vol. 12, No. 2, pp 140 – 158.
[33] Rodrigues, C. K. S. (2006), "Mecanismos de Compartilhamento de Recursos para Aplicações de Mídia Contínua na Internet", UFRJ COPPE/PESC, Rio de Janeiro, Brazil.
[34] De Souza e Silva, E., Figueiredo, D. & Leão, R. (2009), "The Tangram-II Integrated Modelling Environment for Computer Systems and Networks", ACM SIGMETRICS Performance Evaluation Review, Vol. 36, No. 4, pp 45 – 65.

**AUTHORS**

**Ananda Görck Streit** is to receive her B.Sc. degree in Computer Science from the University Center of Brasília (UniCEUB), Brazil. She is applying for a M.Sc. program in the field of Computer Networks and Multimedia Communications. She has published four different research papers on Computer Networks and has coordinated and been actively engaged for three semesters in a research group on Artificial Intelligence. Recently, she has participated in a scholarship program promoted by the brazilian government to study for two semesters in the RWTH Aachen University, Germany.

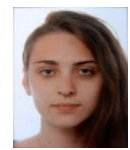

**Carlo Kleber da Silva Rodrigues** received the B.Sc. degree in Electrical Engineering from the Federal University of Paraiba (UFPB) in 1993, the M.Sc. degree in Systems and Computation from the Military Institute of Engineering (IME) in 2000, and the D.Sc. degree in System Engineering and Computation from the Federal University of Rio de Janeiro (UFRJ) in 2006. Currently he is engineer of the Brazilian Army and Professor at the University Center of Brasília (UniCEUB), in Brazil. His research interests include the areas of computer networks, performance evaluation, and multimedia streaming.

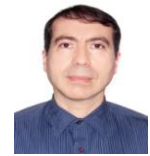